\documentclass[apl,twocolumn]{revtex4-1}%

\usepackage{graphicx}
\usepackage{dcolumn}
\usepackage{bm}

\begin{document}

\preprint{APS/123-QED}

\title{Theoretical expectation of large Seebeck effect in PtAs$_2$ and PtP$_2$}

\author{Kouta Mori$^{1,2}$}
\author{Hidetomo Usui$^{2,3}$}
\author{Hirofumi Sakakibara$^1$}
\author{Kazuhiko Kuroki$^{2,3}$}

\affiliation{$\rm ^1$Department of Engineering Science, The University of Electro-Communications, Chofu, Tokyo 182-8585, Japan}
\affiliation{$\rm ^2$ JST, ALCA, Gobancho, Chiyoda, Tokyo 102-0075, Japan}
\affiliation{$\rm ^3$ Department of Physics, Graduate School of Science, Osaka University, 1-1 Machikaneyama, Toyonaka, Osaka 560-0043, Japan}

\date{\today}

\begin{abstract}
Motivated by a recent observation of good thermoelectric properties in 
PtSb$_2$, we theoretically study related pyrites
in an attempt to 
seek for a material which overcomes the suppression of the Seebeck 
coefficient at high temperatures.
We find that 
PtAs$_2$ and PtP$_2$ are good candidates, where a larger band gap 
than in PtSb$_2$ combined with the  
overall flatness of the band top gives rise to 
a monotonically increasing Seebeck coefficient 
up to high temperatures. This expectation 
has been confirmed quite recently for hole doped PtAs$_2$, 
where a very large power factor 
of $\sim$ 65$\mu$W/cmK$^2$ at $T=440$K is observed.
\end{abstract}

\pacs{  }
\maketitle

Searching good thermoelectric materials is an challenging 
issue both from the viewpoint of fundamental physics and device applications.
From the latter viewpoint in particular, the efficiency of 
thermoelectric materials is characterized by a 
dimensionless figure of merit, $ZT$ with $Z=P/\kappa$, where 
$P=S^2\sigma$ is the power factor, $T$, $S$, $\sigma$ and $\kappa$ 
are the temperature, the Seebeck coefficient, the 
electric conductivity, and the 
thermal conductivity, respectively\cite{Mahanrev}. 
A large power factor 
requires simultaneously large $S$ and $\sigma$, but usually 
materials with large Seebeck coefficient have small conductivity. 
As a general way to overcome this problem, 
one of the present authors along with Arita proposed 
that a band shape that has a flat portion at the top (or the bottom), 
connected to a dispersive portion, can give rise to large 
conductivity and Seebeck coefficient simultaneously. 
This band shape is referred to as the pudding-mold type, and the 
theory was applied to Na$_x$CoO$_2$, a thermoelectric material 
with large metallic conductivity and power factor, 
discovered by Terasaki {\it et al}\cite{Terasaki}.

\begin{figure*}[t]
\includegraphics[width=18cm]{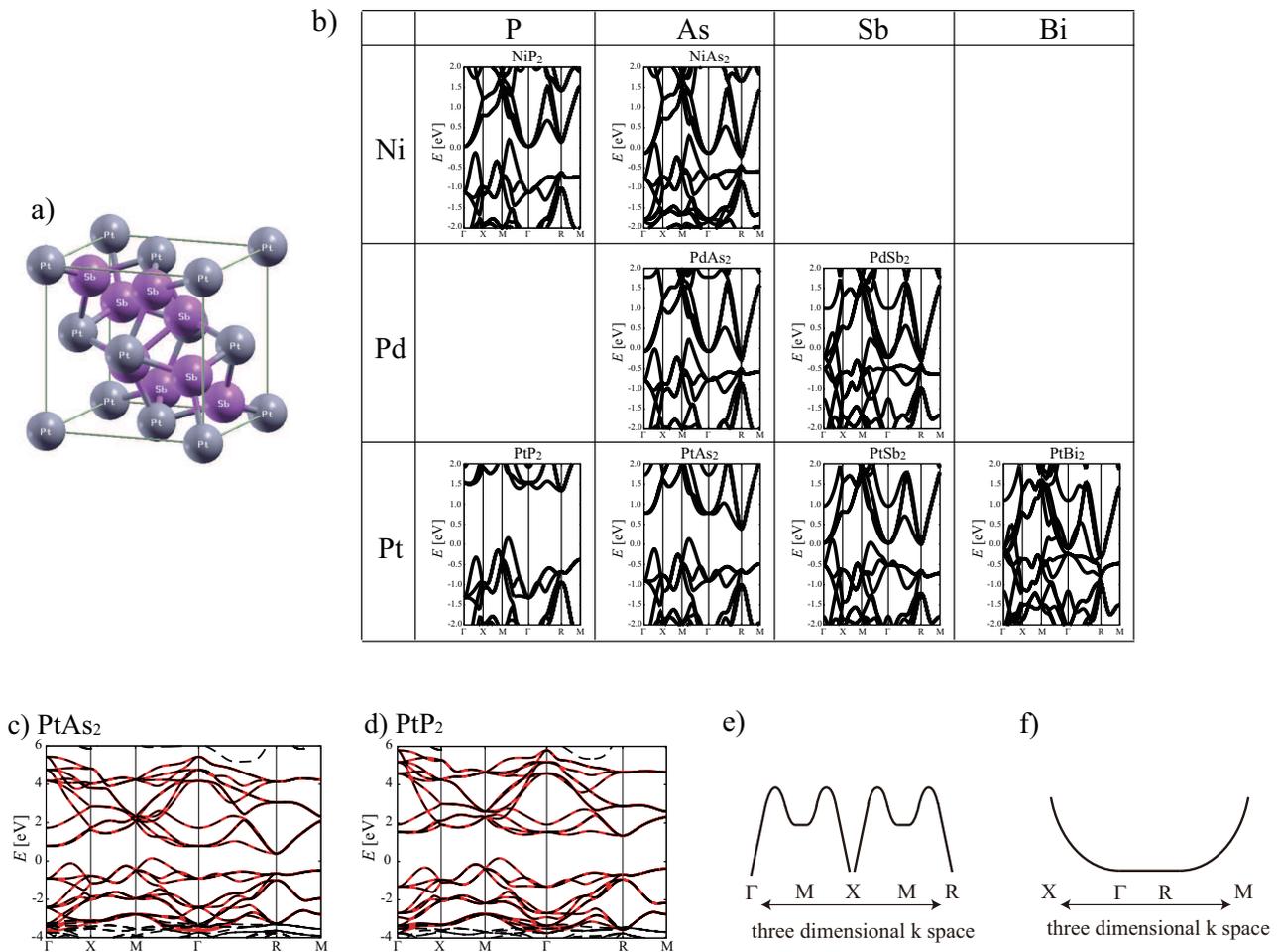}
\caption{
(a) The lattice structure of PtSb$_2$.
(b) Band structures of {\it MPn}$_2$. The blanks indicate 
that experimentally determined lattice structures were not available. 
(c)(d) Tightbinding models of PtAs$_2$ and PtP$_2$.
(e) Schematic figure of the valence band of PtAs$_2$ and PtP$_2$.
(f) Schematic figure of the conduction band of PtP$_2$.
}
\label{fig1}
\end{figure*}

Quite recently, Nishikubo {\it et al.} found that doping holes into 
a cubic pyrite material PtSb$_2$ (Fig.\ref{fig1}(a)) by partially substituting 
Pt by Ir as Pt$_{1-x}$Ir$_x$Sb$_2$ gives rise to a 
metallic conductivity, while exhibiting a large Seebeck effect below 
room temperature, thereby resulting in a large power factor of 
$\sim $40$\mu$W/cmK$^2$ for $x=0.01$\cite{Nishikubo}.
Referring to the band structure calculation in ref.\cite{Emtage}, 
a possible relevance of the peculiar band structure has been pointed out.
Succeedingly, we have analyzed the origin of this large Seebeck effect in 
PtSb$_2$, and concluded that a fairly flat band with dispersive portions,
which we called the ``corrugated flat band'', plays an important 
role\cite{Mori}.
In the study, we also found that the narrowness of the valence-conduction 
band gap results in a 
reduction of the Seebeck coefficient at high temperatures.
We analyzed how the band gap is determined from a tightbinding viewpoint, 
and concluded that 
if we could increase the hopping integral between Pt $5d$ and Sb $4p$ orbitals,
the band gap increases, and the Seebeck coefficient monotonically 
increases up to high temperatures.

In the present study, we theoretically search for related pyrite materials 
that can give a larger band gap and hence a monotonically 
increasing Seebeck coefficient at high temperatures. We  
perform first principles 
band calculation for $MPn_2$ with $M=$Ni, Pd, Pt and 
$Pn=$ P, As, Sb, which 
shows that the band gap for PtAs$_2$ and PtP$_2$ is 
large compared to that of PtSb$_2$. 
By constructing a tightbinding model for the bands near the Fermi level 
exploiting maximally localized Wannier orbitals, 
it is found that 
the large band gap is indeed  due to the large $d$-$p$ hopping. 
Calculating the Seebeck coefficient using the obtained 
tightbinding model, we find a 
monotonically increasing large Seebeck coefficient against 
temperature for these two materials.
For PtP$_2$, we find an ideal band shape in the conduction band structure, 
so that better thermoelectric properties are expected in the electron 
doped regime.

We start with the 
first principles band calculation for cubic pyrites 
$MPn_2$ with $M=$Pt, Pd, Ni and 
$Pn=$ Sb, As, P. We adopt the lattice structure parameter values 
taken from ref.\cite{structure}, and the band structure is calculated 
using the Wien2K package\cite{Wien2k}. 
Here we take $RK_{\rm max}=7$, 512 $k$-points.
The results are summarized in Fig.\ref{fig1}(b). It can be seen that 
in PtAs$_2$ and PtP$_2$, a large band gap opens, while in 
other materials the gap is narrow or absent.

To see the origin of this result in more detail,
we obtain tightbinding models 
exploiting the maximally localized Wannier orbitals\cite{Wannier}.
The bands near the Fermi level consist mainly of the $p$ orbitals, 
but in order to see the $d$-$p$ hybridization explicitly,
we first construct a 44 orbital $d$-$p$ model that considers 
20 (5 orbitals $\times$ 4 sites) {\it Mn} $d$ orbitals 
in addition to the 24 {\it Pn} $p$ orbitals.
The parameters listed in table \ref{table1} are the lattice constant, 
the largest hopping integral $t_{dp}$ between $M$ $d$ and $Pn$ $p$ 
orbitals, the level offset $\Delta E_{dp}$ between $d$ and $p$ orbitals, 
and the band gap $E_g$. 
As analyzed in our previous study\cite{Mori}, 
the band gap opens between the non-bonding (valence bands) 
and the antibonding bands (conduction bands), 
where non-bonding means that the band is almost solely constructed from 
$p$ orbitals.
Therefore, the magnitude of the band gap should roughly 
be proportional to $t_{dp}^2/\Delta E_{dp}$ in the large $\Delta E_{dp}/t_{dp}$ 
limit.

\begin{table*}[!t]
\caption{The lattice constant, the tightbinding parameters, and the band gap 
for the pyrite materials studied in the present paper. All the values are 
normalized by the corresponding values of PtSb$_2$.}
\label{table1}
\begin{tabular}{c| c c |c c |c c c c}
\hline
& NiP$_2$&NiAs$_2$ & PdAs$_2$ & PdSb$_2$& 
PtP$_2$ & PtAs$_2$ & PtSb$_2$ & PtBi$_2$\\  
\hline 
$a$ 				&0.85	&0.89	&0.93	&1.00	&0.88	&0.93	&1.00	&1.04  \\
${\Delta}E_{dp}$	&0.41	&0.33	&1.33	&1.45	&0.63	&0.76	&1.00	&1.08 \\
$t_{dp}$ 			&0.81	&0.72	&0.97	&0.86	&1.22	&1.13	&1.00	&0.88 \\
$E_g$ 			&0.96	&0.87	&0.86	&0.69	&1.46	&1.27	&1.00	&0.81 \\
\hline
\end{tabular}
\end{table*}

There are two competing effects that affect the magnitude of $t_{dp}$;  
one is the unit cell volume (the lattice constant) 
 and the other is the spread of the $p$ or $d$ orbitals.
When Sb is replaced by As or P, the unit cell volume and the $p$ orbital 
spread are both reduced, but the former effect overcomes the latter, thereby 
resulting in an enhanced $t_{dp}$. On the other hand, when 
Pt is replaced by Pd or Ni, the lattice contant barely changes, 
while the $d$ orbital shrinks, giving a smaller $t_{dp}$.
\begin{figure}[t]
\includegraphics[width=8cm]{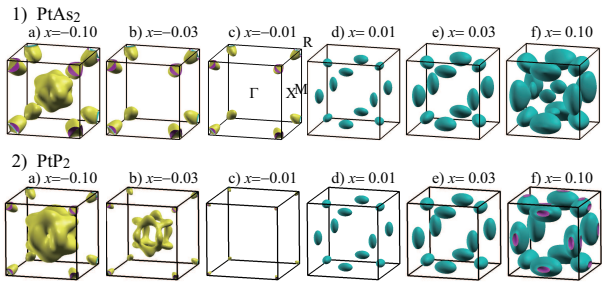}
\caption{Fermi surface of PtAs$_2$ and PtP$_2$. $x$ denotes the doping rate,
where the positive (negative) values imply hole (electron) doping.}
\label{fig2}
\end{figure}   
\begin{figure}[t]
\includegraphics[width=8cm]{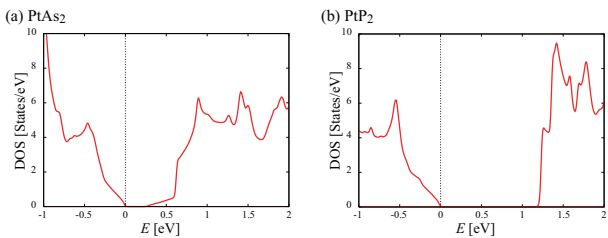}
\caption{The density of states of PtAs$_2$ and PtP$_2$. 
}
\label{fig3}
\end{figure}

Having seen that the band gap of PtAs$_2$ and PtP$_2$ is large 
compared to that of PtSb$_2$, 
we now analyze these two materials in more detail. 
To reproduce the band structure near the Fermi level, 
a 24 orbital model that considers only the $p$ Wannier orbitals 
(eight {\it Pn} per unit cell $\times$ three $p$  
orbitals) suffices.
The band dispersion of the model is shown in fig.\ref{fig1}(c) and (d)
superposed to the original first principles band. 
The Fermi surface of the two materials is obtained using these 
models  for several hole or electron 
doping rates as shown in Fig.\ref{fig2}.
In the hole doped case, it can be seen that the 
Fermi surface pockets are scattered through the 
entire Brillouin zone except around the $\Gamma$ and X points,
indicating that the valence band is essentially 
flat (i.e. have similar energies) 
over a large portion of the zone with some corrugation. This can be seen in the 
band structure in Fig.\ref{fig1}(c),  in which the top positions of the 
valence band at different $k$ points have similar energies except 
around the $\Gamma$, X, and R points. 
Therefore, the band shape can be schematically 
captured as shown in Fig.\ref{fig1}(e). This can be viewed as 
a combination of multiple pudding-mold type bands with a corrugated top,
In the present case, 
the $k$ axis extends through the {\it three dimensional} $k$-space, while  
in previous cases, 
the $k$ axis was either two dimensional $(k_x, k_y)$ like in Na$_x$CoO$_2$ 
or CuAlO$_2$\cite{MoriCuAlO2}, 
or one dimensional as in FeAs$_2$\cite{UsuiFeAs2}.
\begin{figure}[b]
\includegraphics[width=8cm]{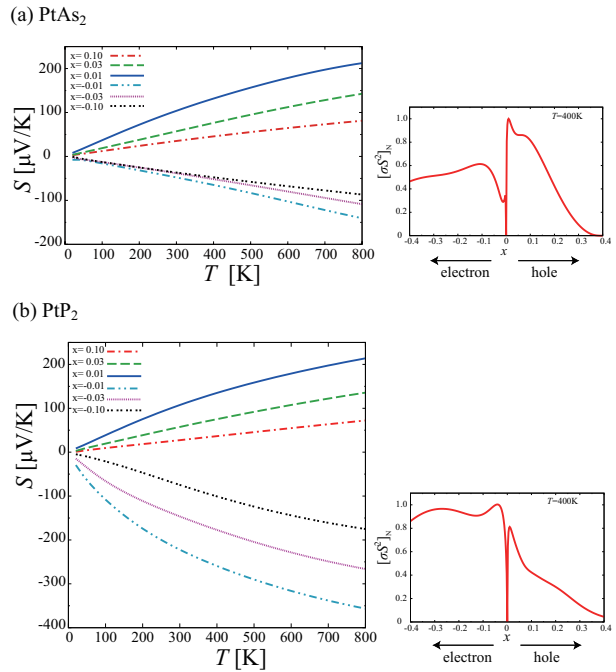}
\caption{(left panels) The calculated Seebeck coefficient 
plotted against temperature for 
PtAs$_2$ and PtP$_2$. $x>0(<0)$ implies hole(electron) doping.
(right) The  power factor against the doping rate at $T=400$K, 
normalized by its maximum value.
}
\label{fig4}
\end{figure}   

On the other hand, in the electron doped case, especially for 
PtP$_2$, the Fermi surface  
is gathered around the $\Gamma$ point meaning that the portion of the 
bands around the zone center is flat  in this case.
As seen in Fig.\ref{fig1}(d), the conduction band of PtP$_2$ 
indeed has a very flat 
bottom that extends over a large portion of the 
three dimensional Brillouin zone, as schematically 
depicted in Fig.\ref{fig1}(f).
This is also clearly seen in the density of states (DOS) of PtP$_2$ shown in 
Fig.\ref{fig3}(b), 
which rises sharply at the conduction band edge, as if the 
material were a 2D system. Thus,  the conduction bands of PtP$_2$ 
have an ideal three dimensional pudding-mold shape.
 
As discussed in ref.\cite{Kuroki}, pudding-mold type band is 
ideal for obtaining good thermoelectric properties, especially the 
power factor. A large thermopower is generally obtained when the 
velocity of electron $(v_e)$ and hole $(v_h)$ 
excitations near the Fermi level have large difference.
When the Fermi level sits close to the band edge, 
the $v_h/v_e$ ratio can in general be large,  
but the absolute values of the 
velocities are usually small and also the Fermi surface is small, 
so that the conductivity is small.
For the pudding-mold type band, on the other hand, 
the large density of states at the top of the band prevents the 
Fermi level from going down rapidly even when a large amount of 
carriers (holes in the present case) is doped. 
This is good for thermopower in that the 
Fermi level stays near the bending point of the band 
even when the Fermi surface volume is large.
When the 
Fermi level sits close to the bending point of the pudding mold type band, 
$v_h/v_e$ ratio is large, resulting in a large Seebeck coefficient. 
At the same time, the conductivity becomes large due to the 
large Fermi surface volume and the dispersive portion (mainly below the 
Fermi level) of the bands.
In this manner, the coexistence of large Seebeck 
coefficient and large conductivity is realized for a wide range of hole 
doping rate.

We now proceed to the calculation of the Seebeck coefficient for 
PtAs$_2$ and PtP$_2$. 
The Seebeck coefficient is calculated within the Boltzmann's equation 
approach using the obtained tightbinding band structure. 
In this approach, 
tensors of the Seebeck coefficient $\bm{S}$ and the 
conductivity  $\bm{\sigma}$ are given as, 
\begin{equation}
\bm{S}=\frac{1}{eT}\bm{K}_1 \bm{K}_0^{-1} \label{Seebeck}
\end{equation}
\begin{equation}
\bm{\sigma}=e^2\bm{K}_0 \label{EC}
\end{equation}
where $e(<0)$ is the elementary charge, $T$ is the temperature、tensors 
$\bm{K}_1$,$\bm{K}_2$ are given as 
\begin{equation}
\bm{K}_n=\sum_{\bm{k}}\tau(\bm{k})\bm{v}(\bm{k})\bm{v}(\bm{k})\left[-\frac{\mathrm{d}f(\epsilon)}{\mathrm{d}\epsilon}(\bm{k})\right]\left(\epsilon(\bm{k})-\mu\right)^n \label{K0K1}.
\end{equation}
Here, $\epsilon(\bm{k})$ is the band dispersion,
$\bm{v}(\bm{k})=\frac{1}{\hbar}\nabla_{\bm{k}}\epsilon(\bm{k})$ is the 
group velocity, $\tau(\bm{k})$ is the quasiparticle lifetime, 
$f(\epsilon)$ is the Fermi distribution function, and $\mu$ is the 
Fermi level (chemical potential).
In the present study, we approximate $\tau$ as a constant, so that 
it cancels out in the Seebeck coefficient. We simply write 
$\sigma_{xx}$ and $S_{xx}$ as $\sigma$ and $S$, respectively.
$\sigma$ and thus the power factor ${\sigma}S^2$ contain the 
constant $\tau$, whose absolute value is not determined.
Therefore, we only discuss the values of the power factor 
normalized by its maximum value as a function of the doping rate.

In Fig.\ref{fig4}, we show the Seebeck coefficient against temperature
for the two materials.
It can be seen that the Seebeck coefficient monotonically increases 
with temperature, in contrast to a similar calculation for 
PtSb$_2$\cite{Mori}, 
and reaches $200\mu$V/K for both PtAs$_2$ and PtP$_2$, 
for the hole doping rate of $x=0.01$ and at temperature $T=$800K.
This is indeed due to the combination of the 
ideal band shape, together with the large band gap. 

We also consider the case of electron doping, since the 
conduction band seems to have an ideal pudding mold type band 
structure for PtP$_2$. The Seebeck coefficient 
for PtP$_2$ indeed 
exceeds $-200\mu$V/K already at 300K, and reaches a very large 
value of $-350\mu$V/K at 800K for $x=-0.01$ ($''-''$ implies electron doping). 
The superiority of the electron doped PtP$_2$ can further be seen in the 
doping dependence of the 
normalized power factor at $T=$400K also shown in Fig.\ref{fig4}. While the 
hole doped regime gives a better power factor for PtAs$_2$, 
in PtP$_2$ the electron doped regime gives larger and more persisting 
power factor compared to that in the hole doped regime.
Based on these results, we expect good thermoelectric properties in 
PtP$_2$ once sufficient amount of electrons is doped.

To summarize, we have studied the band structure and the 
Seebeck coefficient of {\it M}{\it Pn}$_2$ in an attempt to search for 
materials with thermoelectric properties better than PtSb$_2$.
We find that PtAs$_2$ and PtP$_2$ are good candidates with larger
band gap, while Pd and Ni based compounds are not expected to exhibit 
good performance.
Quite recently, the thermoelectric properties of the hole doped 
Pt$_{1-x}$Rh$_x$As$_2$ has been measured experimentally, 
in which a large and monotonically increasing Seebeck coefficient 
at high temperature is found\cite{KudoPtAs2}. 
The combination of this with the metallic behavior of the 
resistivity results in a very large power factor of 
$\sim$ 65$\mu$W/cmK$^2$ at $T=440$K.


We are grateful to M. Nohara and K. Kudo for 
valuable discussion. H.S. acknowledges support from JSPS.

\end{document}